\documentclass[prc,showpacs,superscriptaddress,twocolumn]{revtex4}

\usepackage{amsmath,bm}
\usepackage{graphicx}
\usepackage{epstopdf}
\usepackage{subfigure}
\usepackage{epsfig}
\usepackage{amsmath,amssymb,amsfonts}
\usepackage{color}
\usepackage[utf8]{inputenc}
\usepackage{verbatim}
\usepackage{array}
\graphicspath{{figures/}} 

\begin{document}

\newcommand{\vk}{{\vec k}}
\newcommand{\vK}{{\vec K}}
\newcommand{\vb}{{\vec b}}
\newcommand{{\vp}}{{\vec p}}
\newcommand{{\vq}}{{\vec q}}
\newcommand{\vQ}{{\vec Q}}
\newcommand{\vx}{{\vec x}}
\newcommand{\beq}{\begin{equation}}
\newcommand{\eeq}{\end{equation}}
\newcommand{\half}{{\textstyle \frac{1}{2}}}
\newcommand{\gton}{\stackrel{>}{\sim}}
\newcommand{\lton}{\mathrel{\lower.9ex \hbox{$\stackrel{\displaystyle<}{\sim}$}}}
\newcommand{\ee}{\end{equation}}
\newcommand{\ben}{\begin{enumerate}}
\newcommand{\een}{\end{enumerate}}
\newcommand{\bit}{\begin{itemize}}
\newcommand{\eit}{\end{itemize}}
\newcommand{\bc}{\begin{center}}
\newcommand{\ec}{\end{center}}
\newcommand{\bea}{\begin{eqnarray}}
\newcommand{\eea}{\end{eqnarray}}

\newcommand{\beqar}{\begin{eqnarray}}
\newcommand{\eeqar}[1]{\label{#1} \end{eqnarray}}
\newcommand{\pleft}{\stackrel{\leftarrow}{\partial}}
\newcommand{\pright}{\stackrel{\rightarrow}{\partial}}

\newcommand{\eq}[1]{Eq.~(\ref{#1})}
\newcommand{\fig}[1]{Fig.~\ref{#1}}
\newcommand{\eff}{ef\!f}
\newcommand{\alphas}{\alpha_s}

\renewcommand{\topfraction}{0.85}
\renewcommand{\textfraction}{0.1}
\renewcommand{\floatpagefraction}{0.75}

\title{Probing the in-medium $P_T$-broadening by $\gamma$+HF angular de-correlations}

\author{Sa Wang}
\affiliation{Guangdong Provincial Key Laboratory of Nuclear Science,Institute of Quantum Matter, South China Normal University, Guangzhou 510006, China}
\affiliation{Guangdong-Hong Kong Joint Laboratory of Quantum Matter, Southern Nuclear Science Computing Center, South China Normal University, Guangzhou 510006, China}
\affiliation{Key Laboratory of Quark \& Lepton Physics (MOE) and Institute of Particle Physics, Central China Normal University, Wuhan 430079, China}

\author{Jin-Wen Kang}
\affiliation{Key Laboratory of Quark \& Lepton Physics (MOE) and Institute of Particle Physics, Central China Normal University, Wuhan 430079, China}

\author{Wei Dai}
\affiliation{School of Mathematics and Physics, China University of Geosciences, Wuhan 430074, China}

\author{Ben-Wei Zhang}
\email{bwzhang@mail.ccnu.edu.cn}
\affiliation{Key Laboratory of Quark \& Lepton Physics (MOE) and Institute of Particle Physics, Central China Normal University, Wuhan 430079, China}
\affiliation{Guangdong Provincial Key Laboratory of Nuclear Science,Institute of Quantum Matter, South China Normal University, Guangzhou 510006, China}

\author{Enke Wang}
\affiliation{Guangdong Provincial Key Laboratory of Nuclear Science,Institute of Quantum Matter, South China Normal University, Guangzhou 510006, China}
\affiliation{Guangdong-Hong Kong Joint Laboratory of Quantum Matter, Southern Nuclear Science Computing Center, South China Normal University, Guangzhou 510006, China}
\affiliation{Key Laboratory of Quark \& Lepton Physics (MOE) and Institute of Particle Physics, Central China Normal University, Wuhan 430079, China}

\date{\today}

\begin{abstract}
Angular correlations between vector boson and heavy flavors~(HF) are potentially new effective tools to gain insight into the partonic interactions in the quark-gluon plasma (QGP). In this paper, we present the theoretical study of the azimuthal angular de-correlations of $\gamma+$HF in nucleus-nucleus collisions as a new probe of the in-medium $P_T$-broadening effect. The initial production of $\gamma+$HF in p+p is generated by SHERPA which matches the next-to-leading hard processes with parton shower. The in-medium heavy quark evolution is implemented by a Monte Carlo Langevin simulation, which takes into account the collisional and radiative energy loss. We observe considerable suppression at $\Delta\phi_{\gamma D}\sim\pi$ and enhancement at $\Delta\phi_{\gamma D}<2.8$ in $\gamma+$D azimuthal angular distribution in $0-10\%$ Pb+Pb collisions at $\sqrt{s_{NN}}=$5.02 TeV compared to the p+p baseline, which indicates evident in-medium $P_T$-broadening of charm quarks. We also find that the overall modification patterns of $\gamma+$D angular distribution are sensitive to the selection cut of D meson $p_T$. Furthermore, by constructing the 2D ($x_{J}^{\gamma D}, \Delta\phi_{\gamma D}$) correlation diagram, it's possible to display the respective impact of the two aspects of jet quenching, energy loss and $P_T$-broadening, on the final-state $\gamma+$D observable simultaneously. Additionally, we find weaker angular de-correlations of $\gamma+$B compared to $\gamma+$D which may be helpful to understand the mass hierarchy in heavy-ion collisions. Finally, the nuclear modification of $\Delta \phi_{\gamma D}$ distributions in central $0-10\%$ Au+Au collisions at RHIC energy is provided for completeness.
\end{abstract}

\pacs{13.87.-a; 12.38.Mh; 25.75.-q}

\maketitle

\section{Introduction}
\label{sec:intro}
The strongly-coupled droplet of quark-gluon plasma (QGP) is one of the most intriguing discoveries at the Relativistic Heavy Ion Collider (RHIC) and the Large Hadron Collider (LHC).
The strong interactions between the initial-produced energetic jet and the hot QCD matter, referred as the ``jet quenching" phenomenon, are effective probes of the properties of the QGP, which have been extensively investigated in the past decades~\cite{Gyulassy:2003mc,Qin:2015srf,Vitev:2008rz,CasalderreySolana:2010eh,He:2011pd,Neufeld:2010fj,Senzel:2013dta,Casalderrey-Solana:2014bpa,
Dai:2012am,Milhano:2015mng,Chang:2016gjp,Connors:2017ptx,Zhang:2018urd,Chen:2020pfa}. These studies reveal the two most important aspects of jet quenching in heavy-ion collisions, parton energy loss and transverse momentum broadening, which are closely related to the jet transport coefficient $\hat{q}\equiv d\left\langle p_{\perp}^2 \right\rangle/dL$ quantifying the strength of momentum exchanges transverse to the direction of jet parton caused by the in-medium scattering~\cite{Burke:2013yra,Baier:1996kr,Baier:1996sk,Baier:1998kq,Liu:2015vna,Xie:2019oxg,JETSCAPE:2021ehl,Ru:2019qvz,Kumar:2020wvb}.

Lots of effort has been made to address the in-medium $P_T$-broadening effect in the past few years both on experimental~\cite{Adare:2009vd,Aad:2010bu,Chatrchyan:2012gt,Sirunyan:2017qhf,Sirunyan:2017jic,Adam:2015doa,Adamczyk:2017yhe,Norman:2020grk} and theoretical~\cite{Dominguez:2008vd,DEramo:2012uzl,Wu:2014nca,Mueller:2016gko,Chen:2016vem,Mueller:2016xoc,Luo:2018pto,Ringer:2019rfk,Jia:2019qbl,Blanco:2020uzy,Zakharov:2020sfx,Clayton:2021uuv} sides. Although the measurement of hadron-jet correlations in Au-Au collisions at the RHIC is believed as the indication of the $P_T$-broadening in heavy-ion collisions~\cite{Adamczyk:2017yhe}, it's still a challenge to observe this effect in A+A collisions at the LHC energy~\cite{Ringer:2019rfk}. The Ref.~\cite{Mueller:2016gko} has shown that at the LHC, the in-medium $P_T$-broadening of inclusive dijet in nucleus-nucleus collisions are polluted by the vacuum Sudakov effects therefore hard to be measured in experiment. For other processes such as $Z^0/\gamma$+jet,
 since stronger jet reduction due to energy loss is found at nearside where the multiple jets processes dominate~\cite{Luo:2018pto,Zhang:2018urd}, no significant angular de-correlation of $Z^0/\gamma$+jet is observed in the measurements at LHC~\cite{Chatrchyan:2012gt,Sirunyan:2017qhf,Sirunyan:2017jic}. Hence, there is an urgent need for new observables which are sensitive to the $P_T$-broadening effects.

Heavy flavors are also powerful hard probes to gain insight into the partonic interactions in QGP. In addition to the $R_{AA}$~\cite{Adamczyk:2014uip,Adam:2015sza,Sirunyan:2017xss,Khachatryan:2016ypw,ALICE:2018lyv,Xie:2016iwq} and $v_2$~\cite{Adamczyk:2017xur,Acharya:2017qps,Sirunyan:2017plt} of heavy flavor meson, the recent measurement of the angular correlations between $D^0$ meson and jets by CMS collaboration~\cite{Sirunyan:2019dow} sheds new light on the $P_T$-broadening of heavy quarks due to the in-medium interactions~\cite{Wang:2019xey,Wang:2020bqz,Wang:2020ukj}. The radial distribution of charm quarks in jets is found to broaden to larger radii in Pb+Pb collisions compared to p+p. Yet in fact, even the axis of high $p_T$ jet cannot be treated as a perfect reference to probe the $P_T$-broadening of heavy quarks, since the quenching effects also modify the energy-momentum of jets, the jet axis shifts on the $\eta-\phi$ plane correspondingly. On the other hand, due to the energy loss effect, the events selected in Pb+Pb collisions are actually shifted from higher initial kinematic region than that in p+p, scilicet the ``selection bias''~\cite{Renk:2012ve,Cunqueiro:2021wls}. These issues pose challenges to the studies of the nuclear modification mechanism of the observables which are sensitive to the initial kinematic region.

Therefore, to some extent, the heavy flavors tagged by vector boson ($Z/\gamma$+HF) may be more suitable to study the in-medium $P_T$-broadening. First, the vector bosons do not participate in the strong interaction then are good references to probe the changes of heavy quark momentum. Second, compared to the full jets which are composite of multiple particles, the heavy flavor mesons eliminate the pollution of soft particles in the background to a great extent. It's therefore potentially promising to probe the in-medium $P_T$-broadening effect by investigating the angular de-correlations of $Z/\gamma$+HF in heavy-ion collisions. Third, the same $p_T$ cut for the direct photon would guarantee that the initial kinematics region of A+A events is consistent with that of p+p, which is essential for us to understand the medium modification of the angular correlations of $\gamma+$HF in heavy-ion physics. Additionally, since heavy flavors are hard to be produced in the medium excitation, $\gamma+$HF angular correlations can well exclude the impact from the medium response effects~\cite{Cao:2020wlm}, therefore may be helpful to understand the recent measurement on $Z^0$+hadron~\cite{ATLAS:2020wmg,Sirunyan:2021jkr}.

In this work, we present the first theoretical study of the azimuthal angular correlations between the direct photon ($\gamma$) and heavy flavor hadrons ($\gamma+$HF) in high-energy nuclear collisions. The p+p baseline is provided by the event generator SHERPA~\cite{Gleisberg:2008ta} which computes the next-to-leading order matrix elements matched with parton shower effects~(NLO+PS). We investigate the nuclear modification effects of $\gamma+$D azimuthal angular distributions in central $0-10\%$ Pb+Pb collisions at $\sqrt{s_{NN}}=$5.02 TeV compared to the p+p baseline and find considerable angular de-correlations indicating evident in-medium $P_T$-broadening effects. We will show that the overall modification patterns of $\gamma+$D angular distribution are sensitive to the selection threshold of D meson $p_T$. Furthermore, by constructing the correlation diagram of ($x_{J}^{\gamma D}, \Delta\phi_{\gamma D}$), we provide a chance to display the respective impact of energy loss and $P_T$-broadening effects on the final-state $\gamma+$D observable simultaneously. Additionally, we also estimate the angular de-correlations of $\gamma+$B in A+A collisions, and test the mass effect by comparing with that of $\gamma+$D. At last, we present the prediction of the $\gamma+$D angular de-correlations in central $0-10\%$ Au+Au collisions at $\sqrt{s_{NN}}=$200 GeV.

The remainder of this paper is organized as follows. In Sec.~II, the theoretical frameworks used to study the medium modification of $\gamma+$HF angular correlations would be introduced. In Sec.~III, we will show the main results and give specific discussion on the $\gamma+$HF angular de-correlations. At last, we will summarize this paper in Sec.~IV.

\section{Theoretical framework}
\label{sec:framework}

\begin{figure}[!t]
\begin{center}
\vspace*{0.1in}
\includegraphics[width=3.4in,height=3.1in,angle=0]{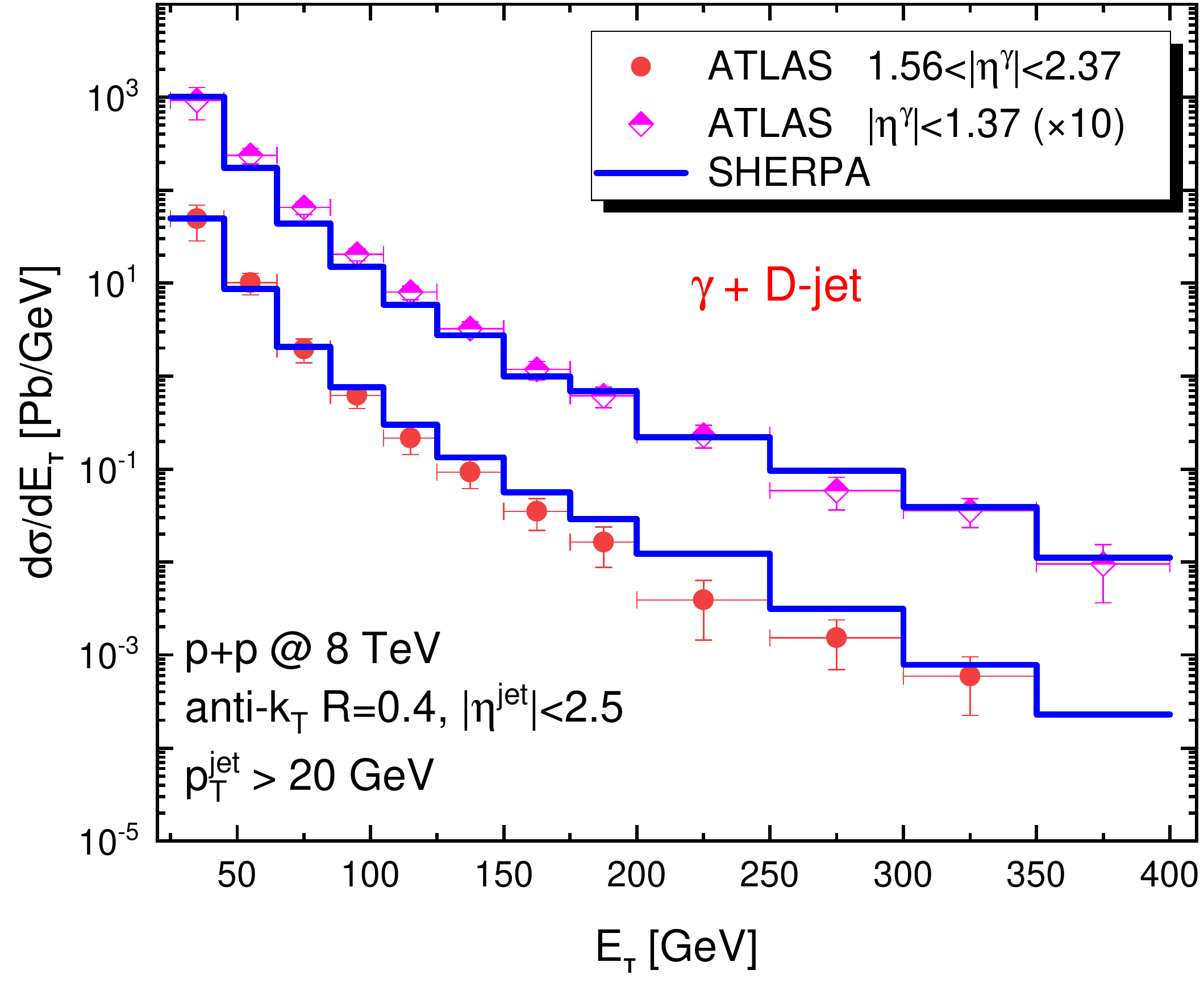}
\includegraphics[width=3.4in,height=3.1in,angle=0]{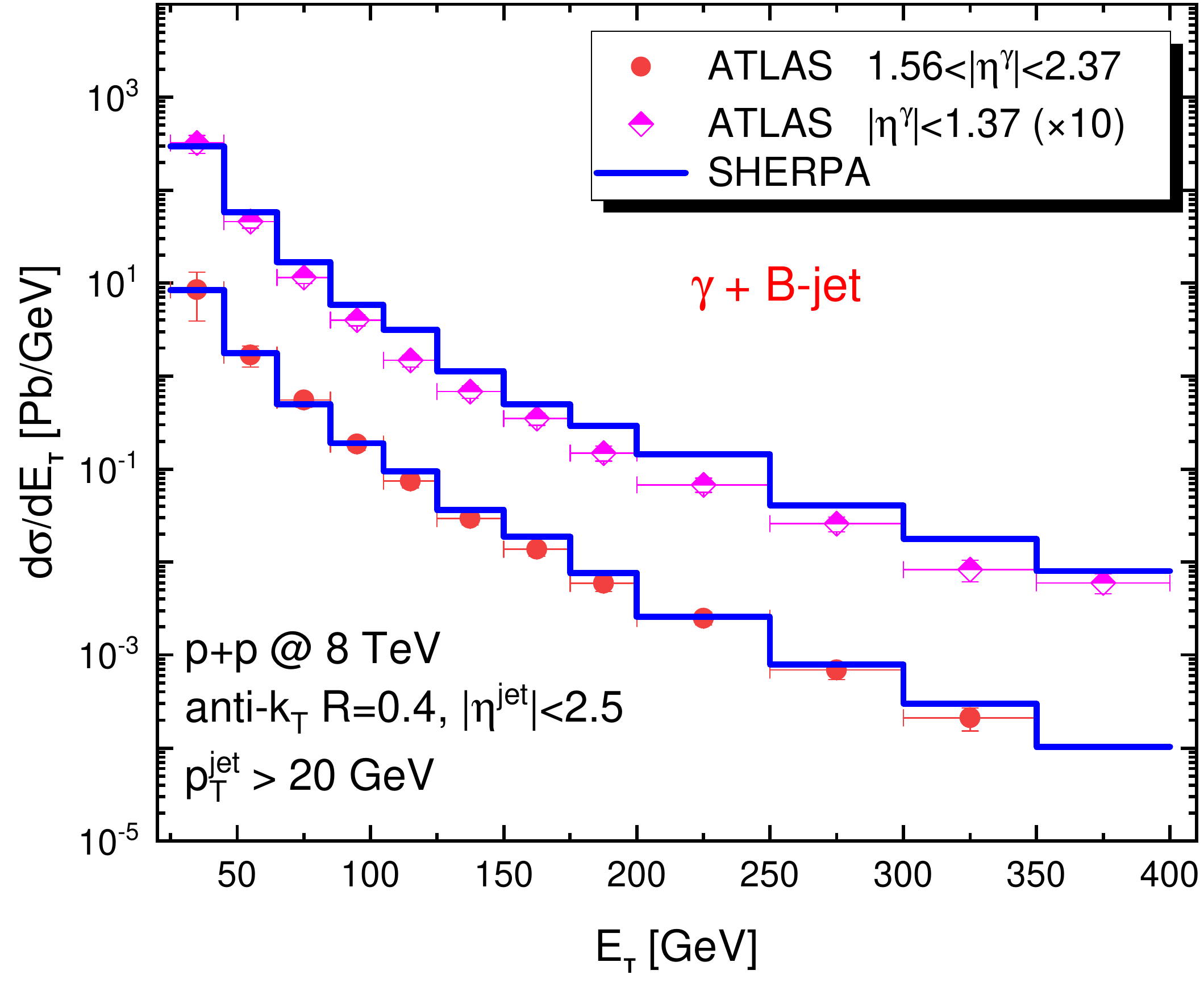}
\vspace*{0.1in}
\caption{Differential cross sections of $\gamma\,+\,$D-jet (upper panel) and $\gamma\,+\,$B-jet (lower panel) versus the transverse energy $E^{\gamma}_T$ of the isolated-photon in p+p collisions at $\sqrt{s}$=8 TeV simulated by SHERPA, compared with ATLAS data~\cite{Aaboud:2017skj}.}
\label{fig:pp-g-cjet}
\end{center}
\end{figure}

In this section, we will first discuss the p+p baseline used in this study, and compare it with the available experimental data. Then we will make a brief introduction of the theoretical framework used to describe the heavy quark evolution in the hot and dense nuclear matter.

In fact, the previous studies~\cite{CMS:2013lua,ATLAS:2016jxf,Zhang:2018urd} had suggested that NLO+PS calculations are prerequisite to investigate the angular correlations between the vector boson and jets. In this work, the production of the $\gamma+$HF in p+p collisions is produced by the Monte Carlo event generator SHERPA-2.2.11~\cite{Gleisberg:2008ta,Krauss:2001iv,Gleisberg:2008fv,Schumann:2007mg}. We use Sherpa in the MC@NLO prescription to generate the photon+jet events containing a photon and two jets, with up to three additional jets, then the photon+HF events can be selected in this photon+jet event sample. The NLO matrix elements are matched with the parton shower using the MC@NLO method~\cite{Frixione:2002ik}, in which the loop diagrams are computed with OpenLoops program~\cite{Buccioni:2019sur}. The NNPDF 3.0 NLO parton distribution function (PDF)~\cite{NNPDF:2014otw} has been chosen in the computation. In Fig.~\ref{fig:pp-g-cjet}, we present the differential cross sections of $\gamma\,+\,$D-jet (upper panel) and $\gamma\,+\,$B-jet (lower panel) as a function of the isolated-photon transverse energy $E^{\gamma}_T$ in p+p collisions at $\sqrt{s}$=8 TeV by computed SHERPA compared to the ATLAS measurements at two $\eta^{\gamma}$ ranges, $|\eta^{\gamma}|<$ 1.37 and 1.56$<|\eta^{\gamma}|<$2.37. Here the final-state jets are reconstructed by Fastjet package~\cite{Cacciari:2011ma} using anti-$k_T$ algorithm~\cite{Cacciari:2008gp} with cone size $R=\sqrt{(\Delta \eta)^2+(\Delta \phi)^2}=$0.4. The D-jets (B-jets) are defined as the jets containing at least one D (B) meson inside the jet cone. The same as the setup used in the ATLAS measurements, the selected D-jets (B-jets) are required to have $p_T^{\rm jet}>$ 20 GeV within $|\eta^{\rm jet}|<$2.5. To select the prompt-photon in the simulations, all candidates are required to pass the Frixione isolation cut, $E^{\rm iso}_T<0.0042\times E^{\gamma}_{T}+4.8$ GeV within a distance R=0.4 around the photon, imposed in the ATLAS measurements~\cite{Aaboud:2017skj}. And we find that the results simulated by SHERPA are consistent with the ATLAS measurements, and a good p+p baseline is the basis of our subsequent study on the angular correlations of $\gamma+$HF in nucleus-nucleus collisions.

Since heavy quarks are viewed as effective hard probes to constrain the transport properties of QGP, a lot of theoretical models~\cite{ vanHees:2007me,CaronHuot:2008uh,Djordjevic:2015hra,He:2014cla,Chien:2015vja,Kang:2016ofv,Cao:2013ita,Alberico:2013bza,Xu:2015bbz,Cao:2016gvr,
Das:2016cwd,Ke:2018tsh,Altenkort:2020fgs,He:2019vgs,Li:2021nim} have been established to confront with the experimental measurements, for reviews see Refs.~\cite{Andronic:2015wma,Cao:2018ews,Dong:2019unq,Dong:2019byy,Cao:2021ces,Zhao:2020jqu}.
In this study, to estimate the nuclear modification effect of the angular correlations of $\gamma+$HF, the initial p+p events produced by SHERPA at parton level are utilized as input of the in-medium evolution within Langevin equations~\cite{Cao:2013ita,Dai:2018mhw,Wang:2019xey,Wang:2020qwe,Wang:2020ukj}.

\begin{eqnarray}
&&\Delta\vec{x}(t)=\frac{\vec{p}(t)}{E}\Delta t\\
&&\Delta\vec{p}(t)=-\Gamma(p,T) \vec{p}\Delta t+\vec{\xi}(t)\sqrt{\Delta t}-\vec{p}_{\rm g}(t)
\label{eq:lang2}
\end{eqnarray}

These two equations represent the position and momentum updates of heavy quarks during the evolution correspondingly. The three terms on the right-hand side of Eq.~(\ref{eq:lang2}) denote the drag term, the thermal stochastic term and the recoil term respectively. $\Gamma$ is the drag coefficient which controls the strength of collisional energy loss of heavy quarks. The thermal stochastic term represents the random kicks suffered on heavy quarks from the thermal quasi-particles in QGP, and obeys a Gaussian distribution with mean value 0 and variance $\kappa$.

\begin{eqnarray}
&& \left \langle \xi^i(t) \right \rangle=0 \\
&& \left \langle \xi^i(t)\xi^j(t') \right \rangle =\kappa \delta^{ij}\delta(t-t')
\label{eq:core}
\end{eqnarray}
The momentum diffusion coefficient $\kappa$ is associated with the drag coefficient $\Gamma$ by the fluctuation-dissipation relation $\kappa=2\Gamma ET=2T^2/D_s$, where $D_s$ denotes the spatial diffusion coefficient. The last negative term -$p_g$ represents the momentum recoil from the radiated gluon caused by the in-medium inelastic interaction, which can be sampled by the gluon spectrum calculated with higher-twist approach~\cite{Guo:2000nz,Zhang:2003yn,Zhang:2003wk,Majumder:2009ge},

\begin{eqnarray}
\frac{dN}{ dxdk^{2}_{\perp}dt}=\frac{2\alpha_{s}C_sP(x)\hat{q}}{\pi k^{4}_{\perp}}\sin^2(\frac{t-t_i}{2\tau_f})(\frac{k^2_{\perp}}{k^2_{\perp}+x^2M^2})^4,
\label{eq:dndxk}
\end{eqnarray}
where $x$ is the energy fraction carried by the radiated gluon and $k_\perp$ the transverse momentum of gluon relative to heavy quarks. $C_s$ is the quadratic Casimir in color representation, and $P(x)$ the splitting function in vacuum~\cite{Wang:2009qb}, $\tau_f=2Ex(1-x)/(k^2_\perp+x^2M^2)$ the gluon formation time. $\hat{q}=q_0(T/T_0)^3p_{\mu}u^{\mu}/E$ is the jet transport parameter~\cite{Chen:2010te}, where $T_0$ is the highest temperature in the most central A+A collisions, and $u^{\mu}$ the velocity of the medium cell where the heavy quark locates. Note that the heavy quark transport coefficients have been extensively investigated in various models, for reviews see Refs.~\cite{Rapp:2018qla,Cao:2018ews}.

In our framework, there are two parameters $\hat{q}_0$ and $D_s$ (or $\kappa$) needed to be determined, and they are usually related by a simple relation $\hat{q}=2\kappa$ approximatively which had been successfully employed to describe the D meson production in nucleus-nucleus collisions~\cite{Cao:2013ita,Li:2019lex,Li:2020kax,Xu:2018gux}. It's noted that in a realistic QGP medium, the two components of $\kappa$ may be different for the relativistic propagation of heavy quarks~\cite{Beraudo:2009pe,Prino:2016cni}, and the $v_Q\sim 1$ may not be a good approximation at low $p_T$. Therefore, we treat $\hat{q}$ and $\kappa$ as two independent parameters to be constrained by experimental data. Note that what matters most in our framework is the simultaneous description on the energy loss both of the light and heavy flavors. Hence firstly, we use the values of $q_0$ extracted base on the identified hadron production in nucleus-nucleus collisions in our precious studies~\cite{Ma:2018swx}, in which $q_0=0.6$~GeV$^2$/fm (RHIC) and $q_0=1.2$~GeV$^2$/fm (LHC) are obtained. After the $q_0$ is fixed, then we extract the best values $D_s(2\pi T)=5$ at the RHIC and $D_s(2\pi T)=4$ at the LHC energy respectively by a $\chi^2$ fitting to the D meson $R_{AA}$ data~\cite{Sirunyan:2017xss,ALICE:2018lyv,Adamczyk:2014uip,Xie:2016iwq}, which are consistent with the Lattice QCD calculation of $D_s(2\pi T)=3.7\sim7$~\cite{Francis:2015daa}.

We assume that the number of the radiated gluon during a time step obeys Possion probability distribution,
\begin{eqnarray}
f(n)=\frac{\lambda^{n}}{n!} e^{-\lambda}
\label{eq:pn}
\end{eqnarray}
where the parameter $\lambda$ denotes the mean number of the radiated gluon and can be calculated by integrating Eq.~(\ref{eq:dndxk}).

\begin{eqnarray}
\lambda(t,\Delta t)=\Delta t\int dxdk^2_{\perp} \frac{dN}{dxdk^2_{\perp} dt}
\label{eq:intdndx}
\end{eqnarray}
During every evolution time step, we first estimate the total probability of inelastic scattering $P_{\rm inel}(t,\Delta t)=1-\lambda e^{-\lambda}$ to determine whether radiation occurs. If it occurs, the radiation number $n$ will be sampled based on Eq.~(\ref{eq:pn}), and the energy-momentum of the radiated gluon can be sampled by the gluon spectrum in Eq.~(\ref{eq:dndxk}) one-by-one. To avoid the divergence at $x\rightarrow 0$, only the gluon with energy above the Debye screening mass $\mu_D=\sqrt{4\pi\alpha_s}T$ is allowed to emit, which ensures that heavy quarks can reach the thermal equilibrium after enough long propagation time. Note that only heavy quarks are taken for the Langevin simulation, as for the light partons in the heavy-flavor jets we consider their radiative energy loss based on the higher-twist formalism, because the medium-induced radiation is the dominant energy loss mechanism for light flavors.

The (2+1)D viscous hydrodynamic model~\cite{Shen:2014vra} has been used to describe the time-space evolution of the expanding QCD fireball. The initial spacial production vertex of $\gamma+$HF in QGP is sampled by Glauber model~\cite{Miller:2007ri}. And we assume that the in-medium evolution stops when the local temperature around the heavy quark is lower than $T_c=165$~MeV. After the in-medium evolution, the fragmentation of heavy quarks into heavy flavor mesons ($c\rightarrow D$ and $b\rightarrow B$) is achieved by the Lund symmetric fragmentation function~\cite{Andersson:1983ia}. It should be noted that the coalescence mechanism has not been considered in this work, which is believed to play an important role in the hadronization of heavy quarks, and it may cause considerable systematic errors for heavy meson spectra at $p_{HQ}<4$ GeV~\cite{Cao:2013ita,Cao:2016gvr,He:2019vgs,Cao:2019iqs}. Hence some further efforts should be made to address the possible effects on the azimuthal angular correlation of $\gamma$+HF from the coalescence hadronization mechanism.

The Langevin transport approach has been applied to study the production of $b\bar{b}$ dijets~\cite{Dai:2018mhw} and $Z^0\,+\,$b-jet~\cite{Wang:2020qwe} in our previous works. During the last two years, it has also been successfully employed to estimate the radial profile of heavy flavor meson in jets in heavy-ion collisions~\cite{Wang:2019xey,Wang:2020bqz,Wang:2020ukj}, and gives decent agreement with the CMS measurements~\cite{Sirunyan:2019dow}.
\section{Results and discussions}
\label{sec:results}
In this section, we investigate the azimuthal angular correlations of $\gamma+$D in high-energy nuclear collisions. In particular, we propose that the $P_T$-broadening of charm quarks due to the in-medium interactions could be reflected in the modification of $\Delta \phi_{\gamma D}=|\phi_{\gamma}-\phi_{D}|$ distribution in heavy-ion collisions. And we find that the two aspects of jet quenching effect, namely energy loss and $P_T$-broadening, can be simultaneously displayed in the ($\Delta \phi_{\gamma D}, x_J^{\gamma D}$) correlation diagram, where $x_J^{\gamma D}=p_T^D/p_T^{\gamma}$. Additionally, we predict that, by comparing the medium modifications of $\Delta \phi_{\gamma D}$ and $\Delta \phi_{\gamma B}$ distributions in central Pb+Pb collisions, the mass effect of jet quenching between charm and bottom quarks can be well addressed. At last, we present the calculations of $\gamma+$D angular correlations in p+p and Au+Au collisions at RHIC energy.

\begin{figure}[!t]
\begin{center}
\vspace*{0.1in}
\includegraphics[width=3.4in,height=3.8in,angle=0]{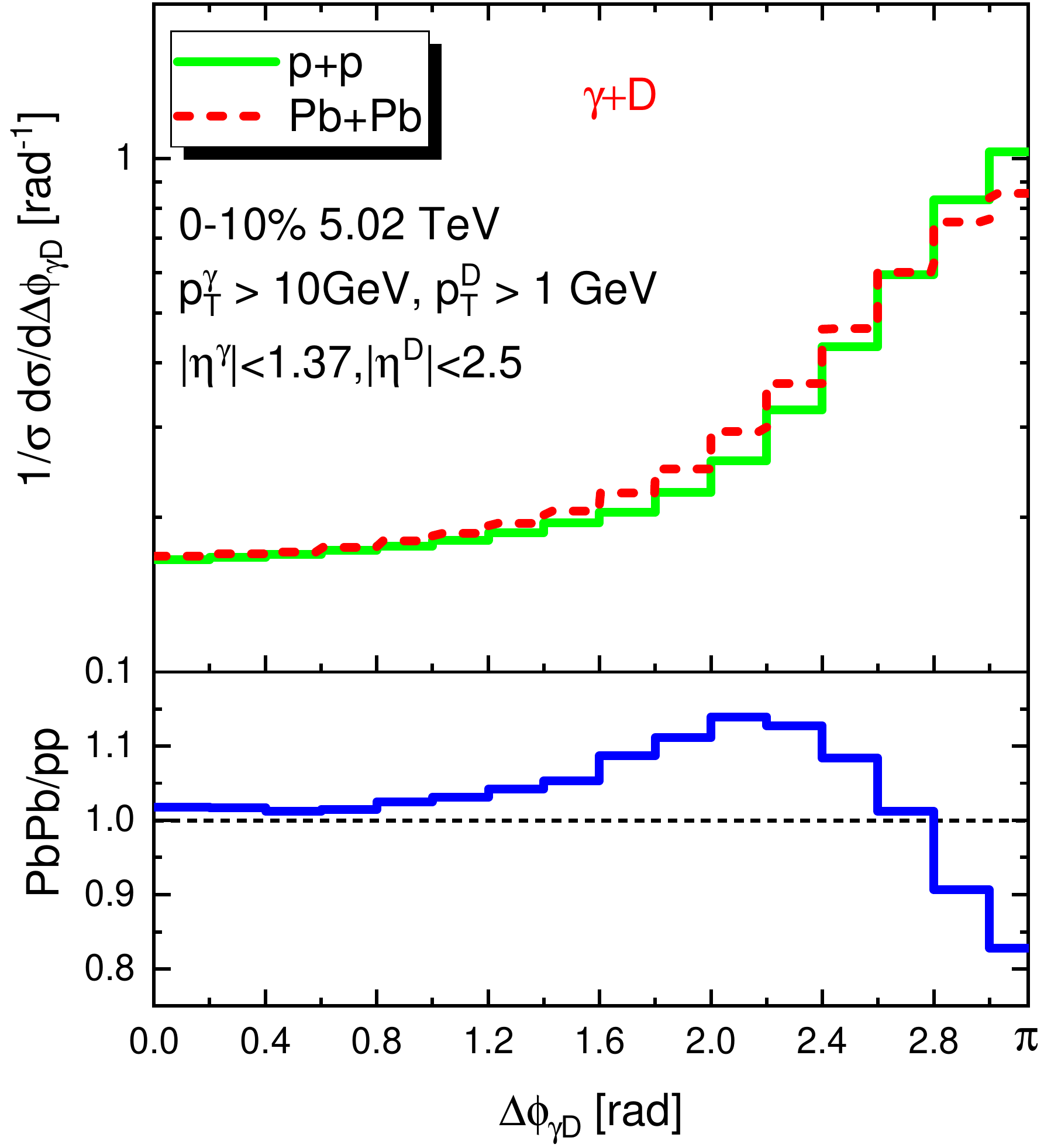}
\vspace*{0.1in}
\caption{The normalized distribution of the azimuthal angular difference ($\Delta \phi_{\gamma D}=|\phi_{\gamma}-\phi_{D}|$) between the isolated-photon and D meson in p+p and 0-10\% Pb+Pb collisions at $\sqrt{s_{NN}}$=5.02 TeV. The ratio of the normalized distributions in Pb+Pb to that in p+p is also plotted in the lower panel.}
\label{fig:ppAAphi}
\end{center}
\end{figure}

In Fig.~\ref{fig:ppAAphi}, we show the normalized distributions of azimuthal angular difference ($\Delta \phi_{\gamma D}=|\phi_{\gamma}-\phi_{D}|$) between the isolated-photon and D meson in p+p and central 0-10$\%$ Pb+Pb collisions at $\sqrt{s_{NN}}$=5.02 TeV, as well as their ratio (PbPb/pp) in the lower panel. It's noted that the requirement  $p_T^{\gamma}>10$~GeV can well constrain the initial kinematics of the selected $\gamma+$D events in Pb+Pb collisions to be consistent with that in the p+p baseline, which makes it available to compare the quenched events with their initial distributions. All the selected D mesons are required to have $p_T^D>$ 1 GeV due to the detector resolution at the LHC, which may be used to suppress the contamination of the background in experiment. In the upper panel of Fig.~\ref{fig:ppAAphi}, in Pb+Pb collisions we observe distinct reduction of $\gamma+$D events at the back-to-back region ($\Delta\phi_{\gamma D}\sim \pi$) in their normalized azimuthal angle correlation distribution compared to the p+p baseline. Of course, it's easier to see the medium modification by illustrating the ratio (PbPb/pp) in the bottom panel of of Fig.~\ref{fig:ppAAphi}. We find considerable suppression of the ratio ($\sim 0.82$) at $\Delta\phi_{\gamma D}\sim\pi$ and enhancement (maximum value $\sim$ 1.15) at $\Delta\phi_{\gamma D}\sim 2.1$, and the modification trends to be invisible at $\Delta\phi_{\gamma D}<1.2$. It no doubt indicates the angular de-correlations between the direct photon and charm quarks in nucleus-nucleus collisions due to the strong in-medium interactions of charm quarks.

\begin{figure}[!t]
\begin{center}
\vspace*{0.1in}
\includegraphics[width=3.4in,height=2.6in,angle=0]{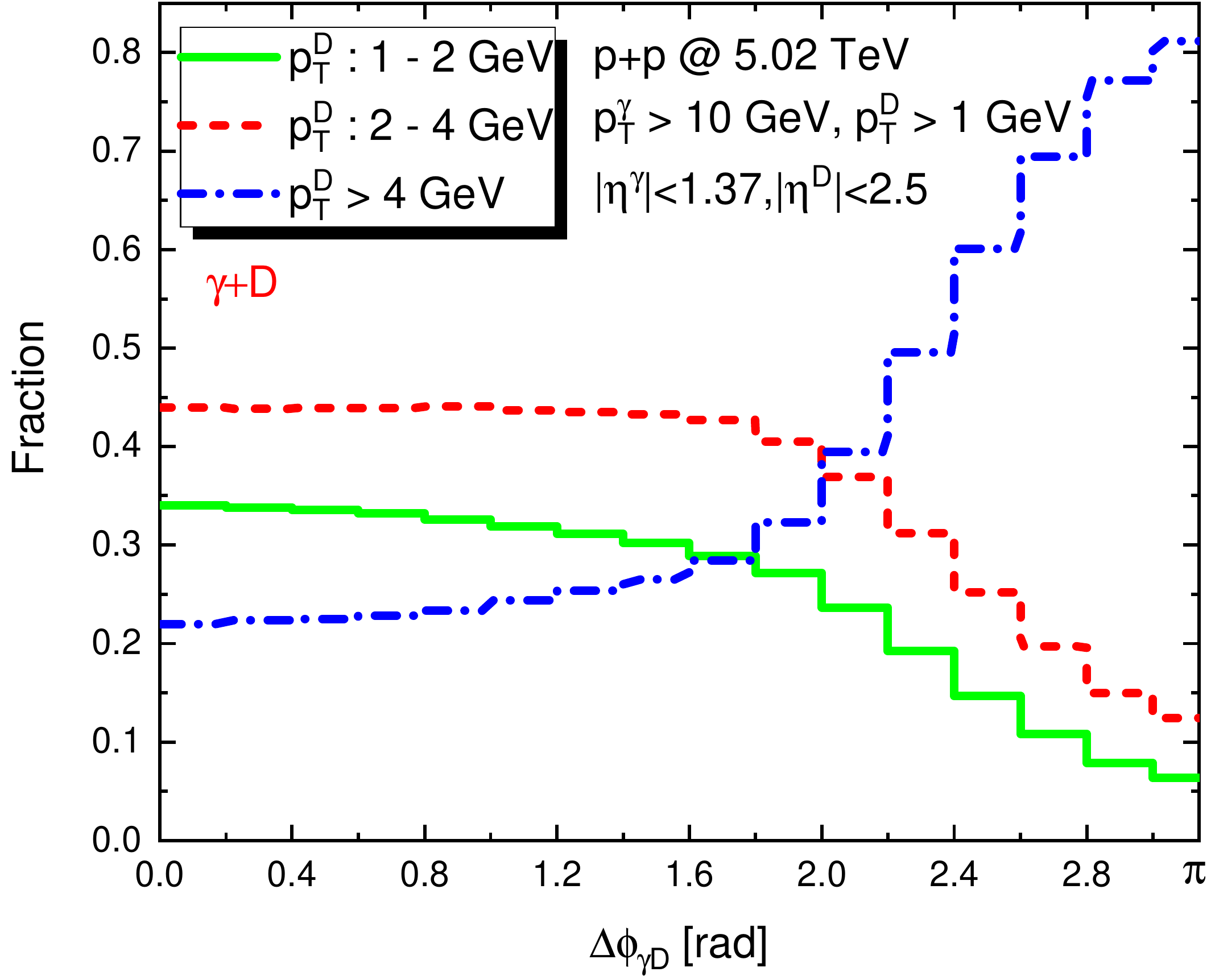}
\includegraphics[width=3.4in,height=2.6in,angle=0]{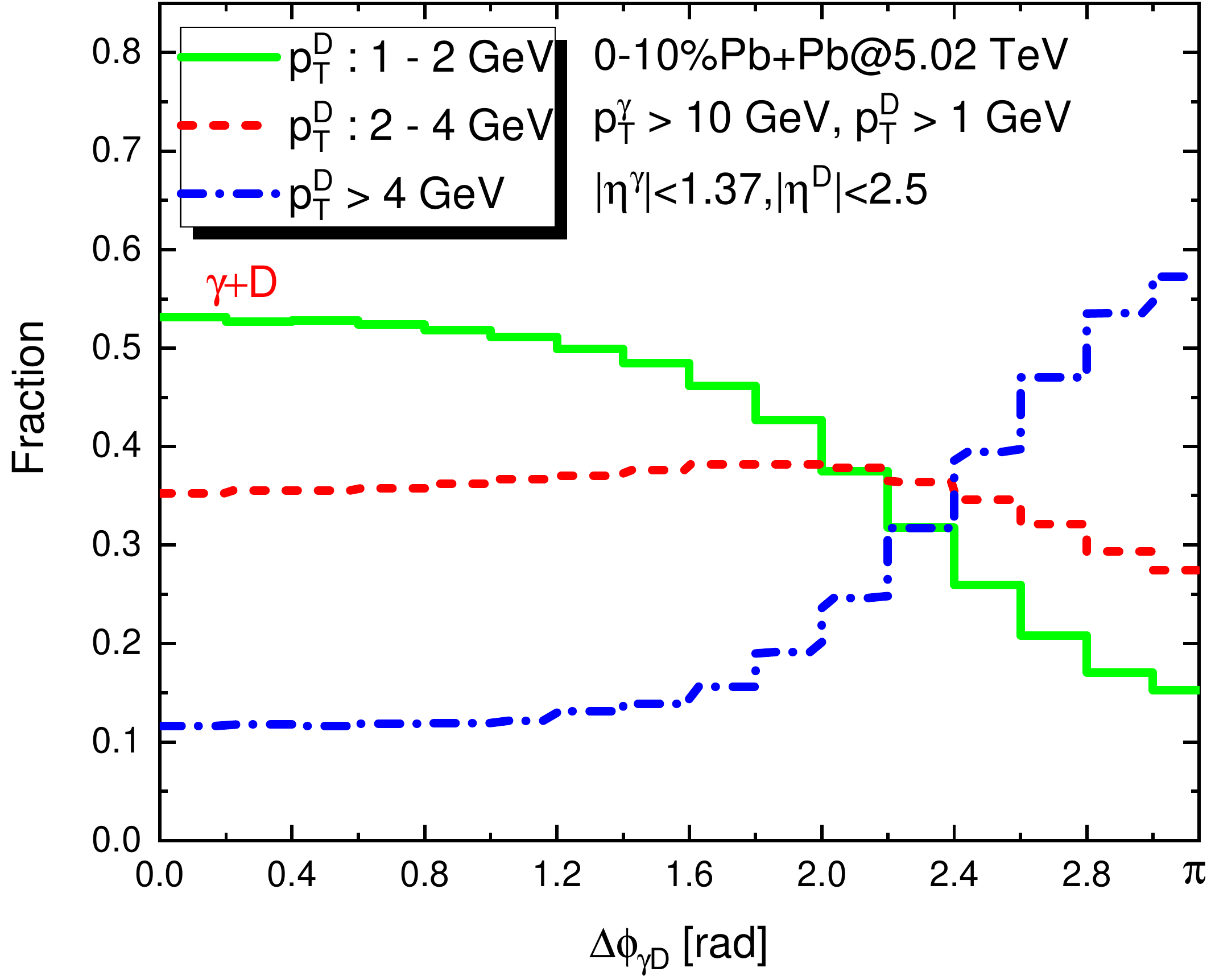}
\vspace*{0.1in}
\caption{The fractional contributions in the total $\Delta \phi_{\gamma D}$ distribution from different D meson $p_T$ ranges in p+p (upper panel) and 0-10$\%$ Pb+Pb (lower panel) collisions at $\sqrt{s_{NN}}$=5.02 TeV.}
\label{fig:fraction}
\end{center}
\end{figure}

To further understand the angular de-correlations of $\gamma+$D in nucleus+nucleus collisions, it's always essential and helpful to analyse the contributions in the $\Delta\phi_{\gamma D}$ distribution from different D meson $p_T$ bins. As shown in Fig.~\ref{fig:fraction}, we estimate the fractional contributions of D mesons with $1<p_T^D<2$ GeV, $2<p_T^D<4$ GeV and $p_T^D>4$ GeV to the total $\Delta\phi_{\gamma D}$ distribution both in p+p (upper panel) and $0-10\%$ Pb+Pb (lower panel) collisions, where the $\rm Fraction(\Delta\phi)$ is calculated as follows,

\begin{eqnarray}
\rm Fraction(\Delta\phi)|_{(p_T^{\rm min},p_T^{\rm max})}&=&\frac{\frac{dN}{d\Delta\phi}|_{p_T^{\rm min}<p_T^D<p_T^{\rm max}}}{\frac{dN}{d\Delta\phi}|_{p_T^D>1}}
\end{eqnarray}

\begin{figure}[!t]
\begin{center}
\vspace*{0.1in}
\includegraphics[width=3.4in,height=3.1in,angle=0]{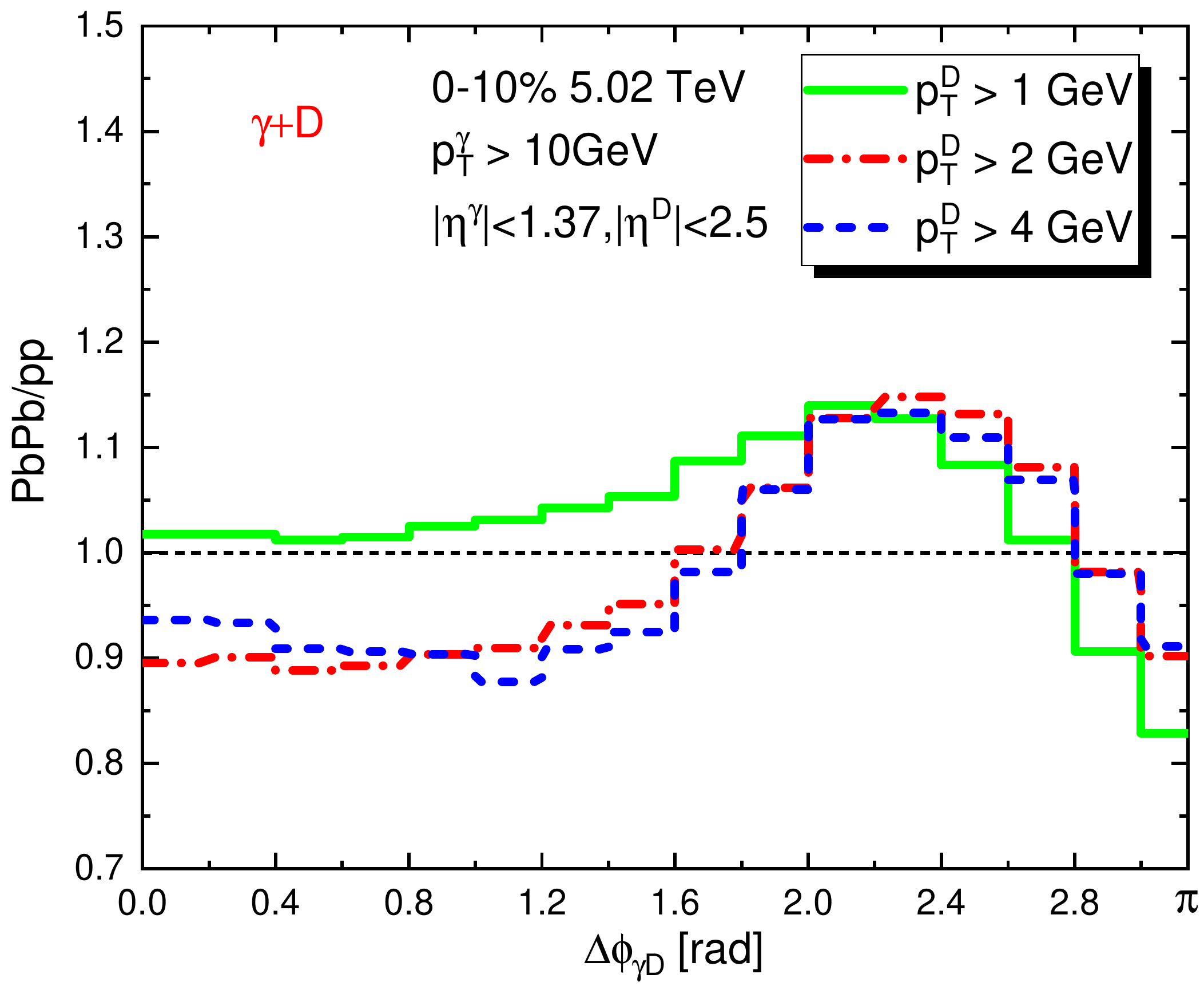}
\includegraphics[width=3.4in,height=3.1in,angle=0]{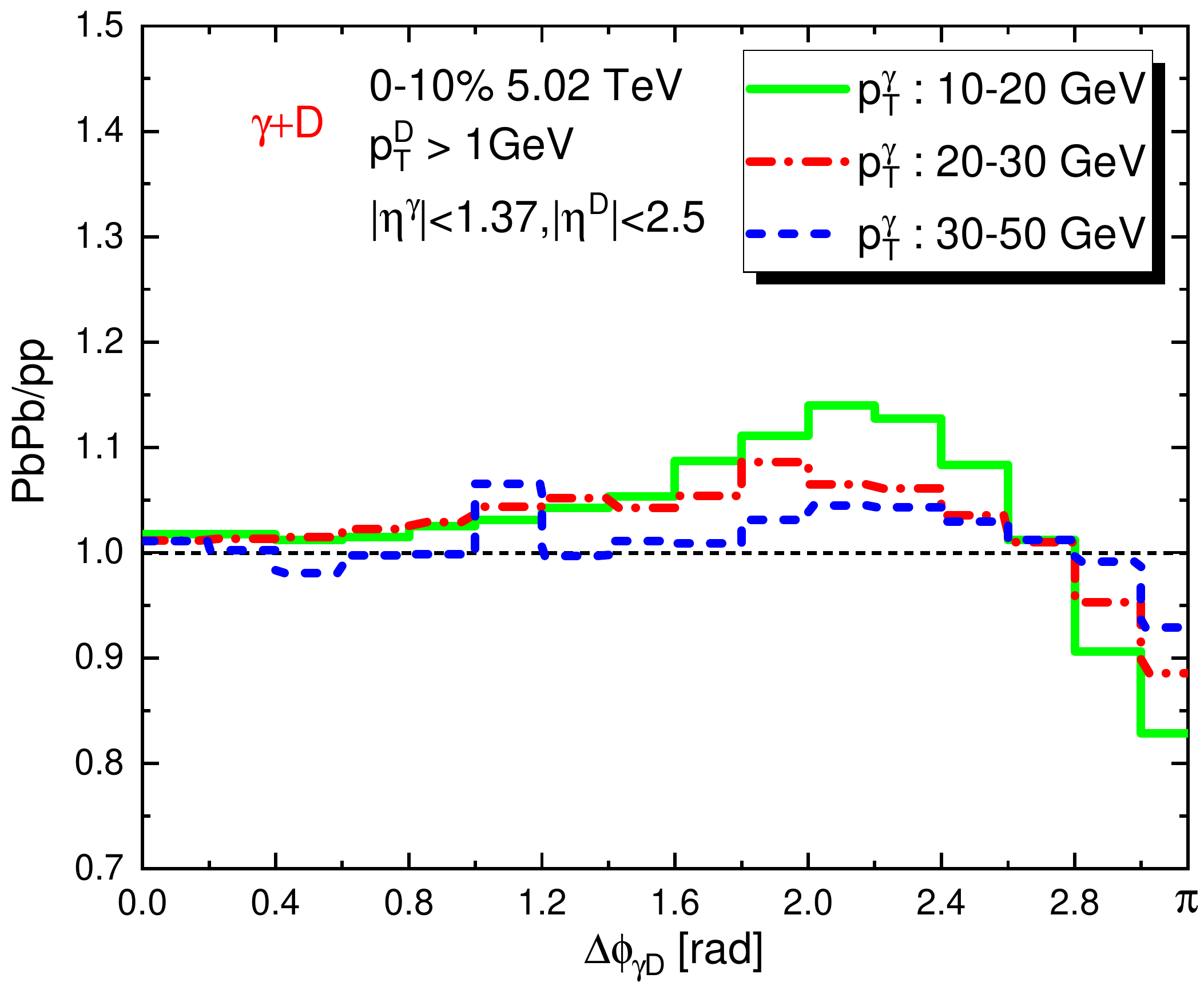}
\vspace*{0.1in}
\caption{Ratios of the normalized $\Delta \phi_{\gamma D}$ distribution of 0-10$\%$ Pb+Pb collisions to p+p, for different $p_T^D$ cut (upper panel) and different $p_T^{\gamma}$ cut (lower panel).}
\label{fig:ptDcut}
\end{center}
\end{figure}

In p+p collisions, we observe that the higher $p_T$ ($p_T^D>4$ GeV) D mesons dominate the large azimuthal angle ($\Delta\phi_{\gamma D}>3\pi/4$) region. In Pb+Pb collisions, we find a significant reduced contribution from higher $p_T$ ($p_T^D>4$ GeV) D mesons due to the energy loss of charm quarks and respectively the enhanced contributions from lower $p_T$ D mesons. At $\Delta\phi_{\gamma D}<2.0$, lower $p_T$ ($p_T^D<4$ GeV) D mesons are the dominant contribution both in p+p and Pb+Pb, which implies that the overall medium modification pattern of $\Delta\phi_{\gamma D}$ distribution may be sensitive to the kinematic cut in event selection at this region. Accordingly, in the upper panel of Fig.~\ref{fig:ptDcut}, we present the calculated medium modification of $\Delta\phi_{\gamma D}$ distribution in $0-10\%$ Pb+Pb collisions for different $p_T^D$ cut. We find that, as D meson selection cut increases, the enhancement at $\Delta\phi_{\gamma D}<2.0$ gradually disappears and turns into suppression. It can be explained by the fact that it is easier to deflect lower $p_T$ heavy quarks into large angles than deflect higher $p_T$ ones. In the lower panel of Fig.~\ref{fig:ptDcut}, the medium modifications of $\Delta\phi_{\gamma D}$ distribution are also estimated within three $p_T$ windows of the direct photon. We find that the medium modifications of $\Delta\phi_{\gamma D}$ gradually decrease with the enhancement of $p_T^{\gamma}$ ranges, but still clear suppressions at $\Delta\phi_{\gamma D}\sim \pi$ can be observed. This $p_T^{\gamma}$ dependence of medium modification can be understood as follows, as $p_T^{\gamma}$ range increases, the initial energy of the correlated charm quarks is also enhanced, however we know that in-medium scattering has weaker influence on the more energic charm quarks. These discussions and estimations may be useful for the future experimental measurements on the medium modification of $\gamma+$D angular de-correlations at the LHC.

\begin{figure}[!t]
\begin{center}
\includegraphics[width=3.7in,height=3.1in,angle=0]{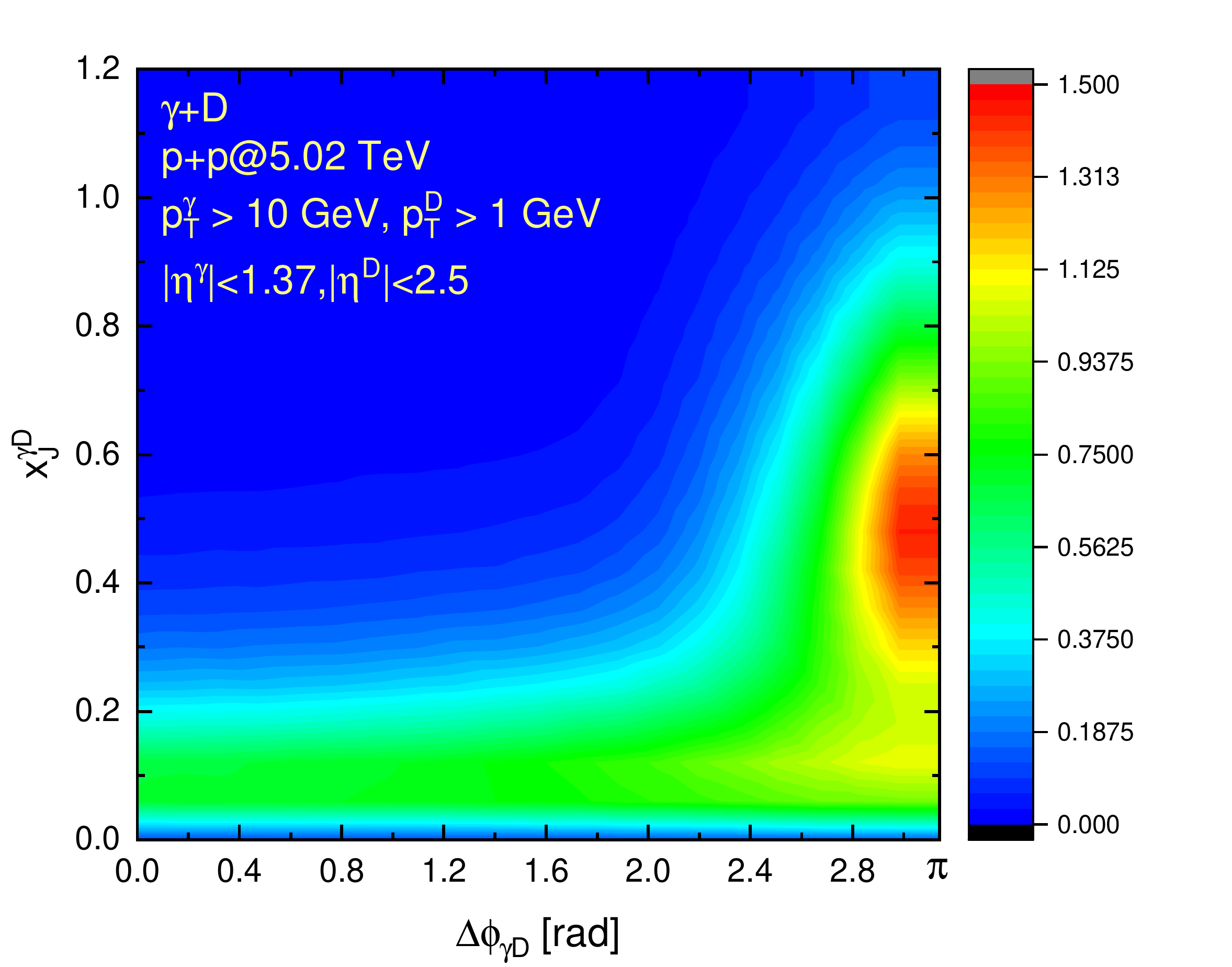}
\includegraphics[width=3.7in,height=3.1in,angle=0]{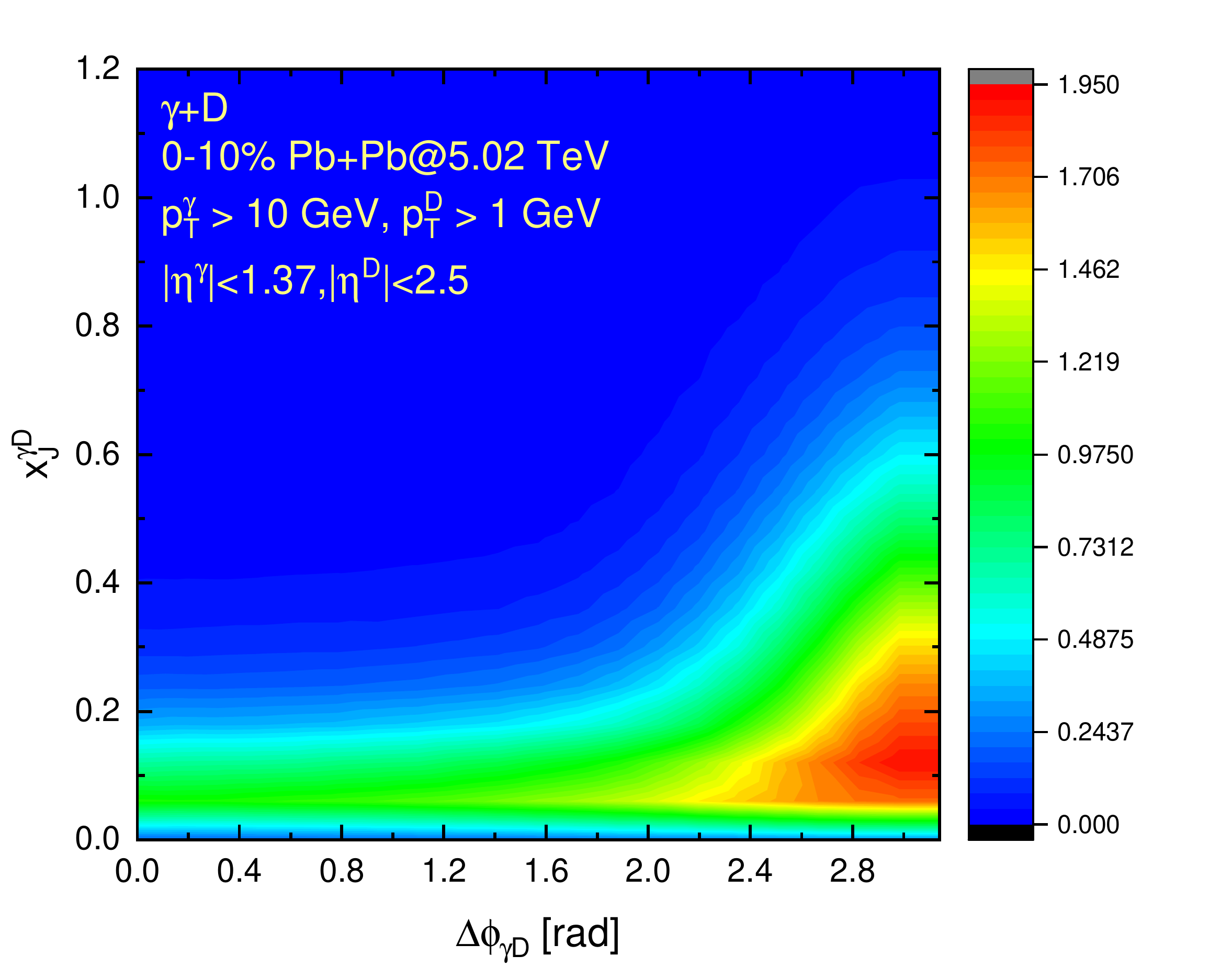}
\vspace*{0.1in}
\caption{2D correlations diagram of $\Delta \phi_{\gamma D}=|\phi_{\gamma}-\phi_{D}|$ and $x_J=p_{T}^D/p_{T}^{\gamma}$ in p+p (upper panel) and 0-10$\%$ Pb+Pb (lower panel) collisions at $\sqrt{s_{NN}}$=5.02 TeV.}
\label{fig:phixj}
\end{center}
\end{figure}

From the above discussions, we find that energy loss and $P_T$-broadening effects have intricate interplay on the medium modification of $\gamma+$D angular correlations. To show their respective impact on the final-state $\gamma+$D observable graphically and clearly, we construct the correlations between $x_{J}^{\gamma D}$ and $\Delta\phi_{\gamma D}$ both in p+p and $0-10\%$ Pb+Pb collisions at $\sqrt{s_{NN}}=$5.02 TeV, as shown in Fig.~\ref{fig:phixj}, where $x_J^{\gamma D}=p_T^D/p_T^{\gamma}$ is the transverse momentum balance between D meson and the direct photon. By comparing the diagrams in p+p and Pb+Pb collisions, we can observe two obvious variations. First, we find the events are concentrated at $x_{J}^{\gamma D}\in[0.3,0.6]$ in p+p collisions but at $x_{J}^{\gamma D}\in[0.05,0.3]$ in Pb+Pb collisions. The shift of $x_{J}^{\gamma D}$ from lager to smaller value is due to the energy loss of charm quarks when they traverse the QGP. Second, we also observe that the highlight region versus $\Delta\phi_{\gamma D}$ is obviously broadened in Pb+Pb collisions compared to p+p, and the correlations at $2.0<\Delta\phi_{\gamma D}<2.8$ are significantly strengthened, which all indicate $\gamma+$D events shift towards smaller $\Delta\phi_{\gamma D}$. The shift of $\Delta\phi_{\gamma D}$ in the ($x_{J}^{\gamma D}$, $\Delta\phi_{\gamma D}$) correlation diagram is no doubt a clear signal of the $P_T$-broadening effect of charm quarks due to the in-medium scattering. Therefore, we propose that, in such an investigation of the correlations between $x_{J}^{\gamma D}$ and $\Delta\phi_{\gamma D}$, the two aspects of jet quenching effect, energy loss and $P_T$-broadening, can be well captured and exhibited simultaneously.

\begin{figure}[!t]
\begin{center}
\vspace*{0.1in}
\includegraphics[width=3.4in,height=2.8in,angle=0]{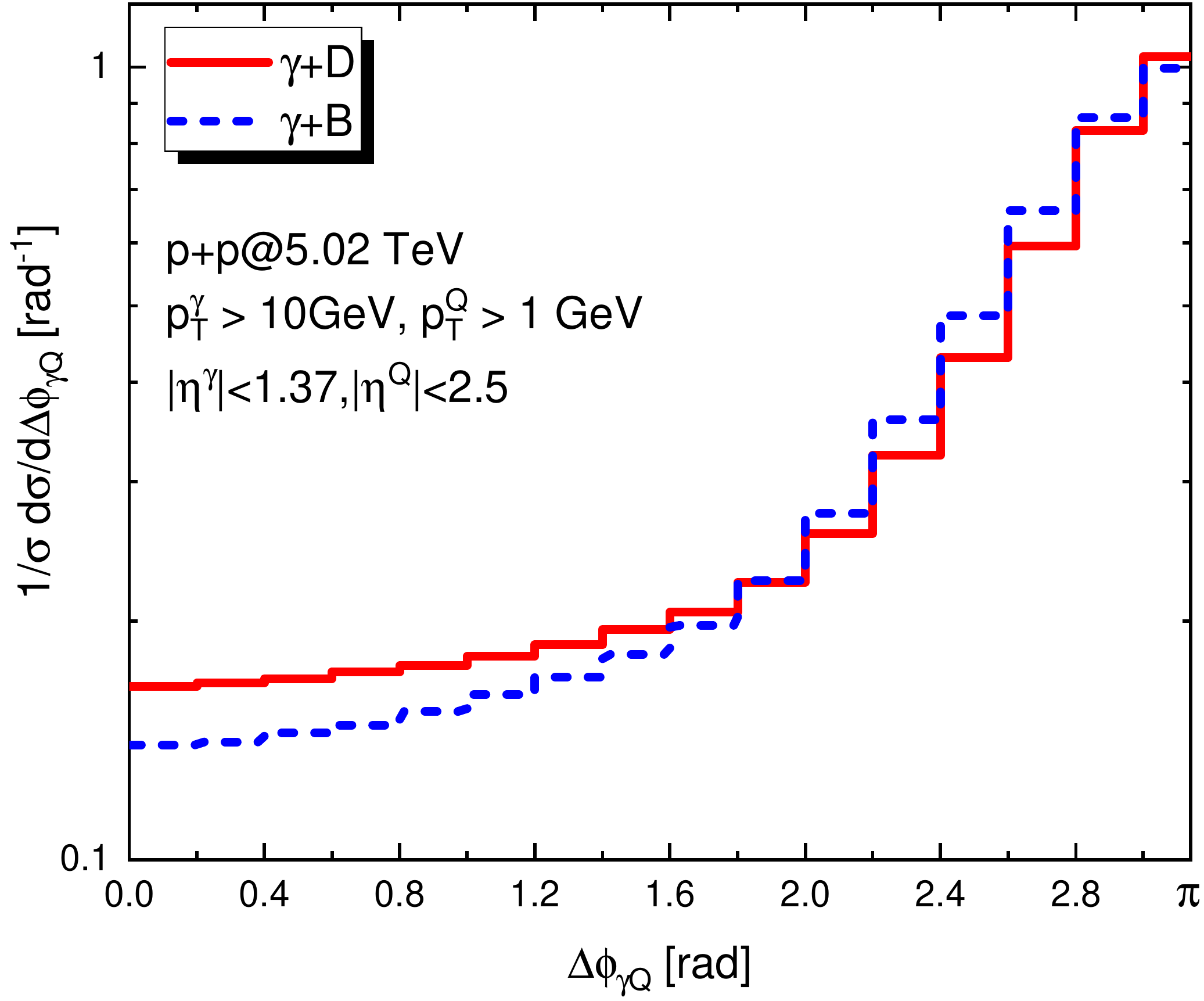}
\caption{The initial $\Delta\phi_{\gamma D}$ and $\Delta\phi_{\gamma B}$ distributions in p+p collisions at 5.02 TeV.}
\label{fig:gDB-pp}
\end{center}
\end{figure}

\begin{figure}[!t]
\begin{center}
\vspace*{0.1in}
\includegraphics[width=3.4in,height=3in,angle=0]{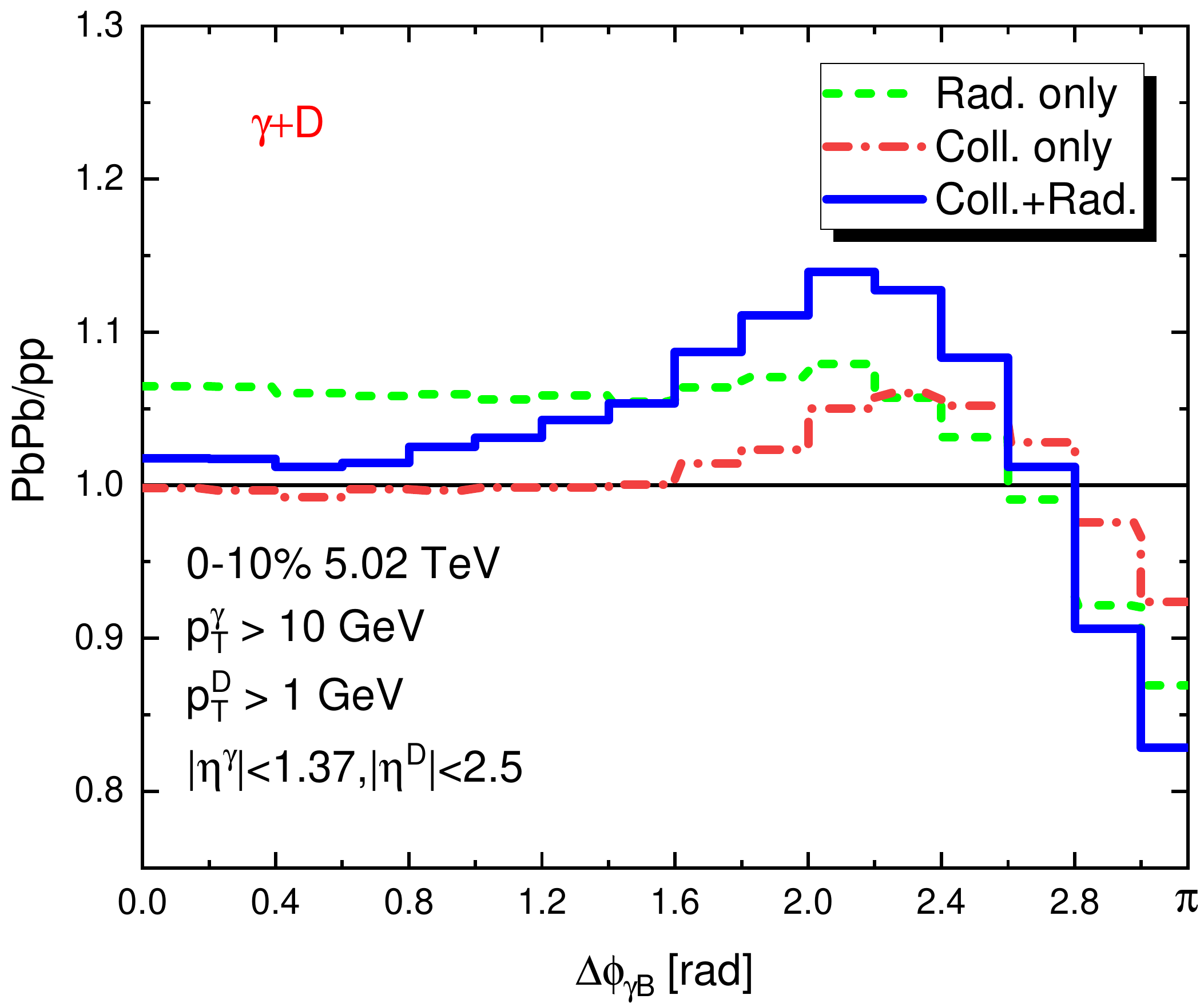}
\includegraphics[width=3.4in,height=3in,angle=0]{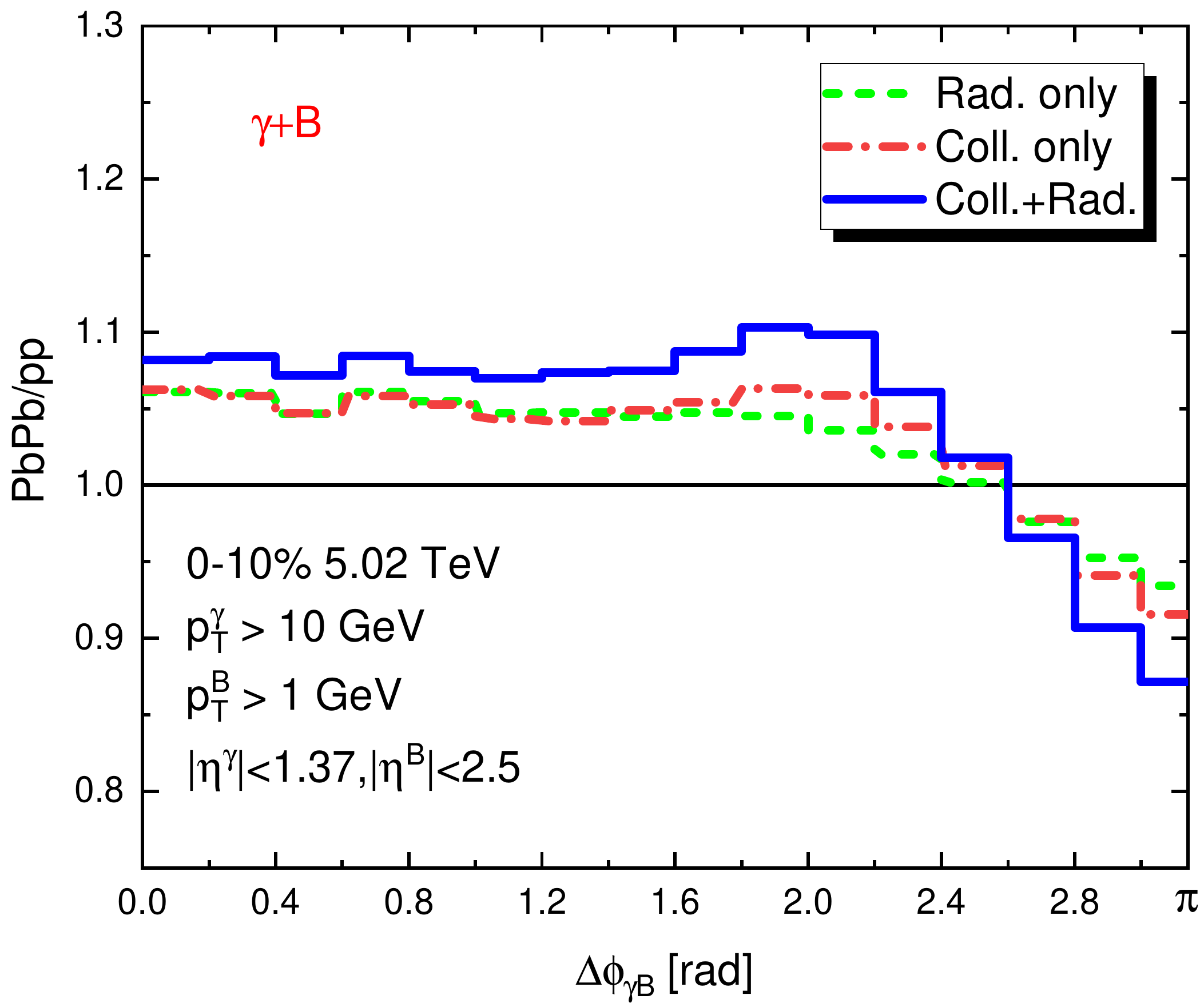}
\vspace*{0.1in}
\caption{Upper panel: ratios of the normalized $\Delta \phi_{\gamma D}$ distribution of 0-10$\%$ Pb+Pb collisions to p+p. Lower panel: ratios of the normalized $\Delta \phi_{\gamma B}$ distribution of 0-10$\%$ Pb+Pb collisions to p+p. Three situations are considered both for $\gamma+$D and $\gamma+$B: only radiative contribution (green dash-line), only collisional contribution (red dot-dash-line), total Rad.+Coll. contributions (blue solid-line).}
\label{fig:ratio-phi-cb}
\end{center}
\end{figure}

The flavor dependence of jet quenching is also an important and interesting topic in heavy-ion physics \cite{Andronic:2015wma,Cao:2018ews,Dong:2019unq,Dong:2019byy,Cao:2021ces,Zhao:2020jqu}. Due to the larger mass, bottom quark was believed to lose less energy than charm quark in the QGP medium, and some indications had been observed in experiment by comparing the nuclear modification factor $R_{AA}$ of D meson and non-prompt $J/\psi$ \cite{Sirunyan:2017xss,Khachatryan:2016ypw}, as well as the most recent reported electron $R_{AA}$ decayed by charm and bottom~\cite{STAR:2021uzu,PHENIX:2022wim}. We have also noted that some other observables may provide new perspectives to gain insight into the mass hierarchy of jet quenching~\cite{Li:2017wwc,Wang:2020ukj}. In this study, for $\gamma+$HF, a good correspondence can be established in p+p and A+A events by using the same $p_T^{\gamma}$ selection cut, which makes it available to compare the strength of the in-medium $P_T$-broadening of charm and bottom quarks straightforward.

It's of necessity and interest to compare the initial azimuthal angular distributions of $\gamma$+D and $\gamma$+B in p+p before proceeding into the discussions in A+A collisions. As shown in Fig.~\ref{fig:gDB-pp}, we plot the normalized distributions of $\Delta\phi_{\gamma D}$ and $\Delta\phi_{\gamma B}$ in p+p collisions at 5.02 TeV. We observe that $\gamma$+B shows steeper angular distribution compared to $\gamma$+D especially at the region of $\Delta\phi<\pi/2$. The initial deference between $\gamma$+D and $\gamma$+B in p+p spectra may also have influence on the medium modification to be observed in A+A collisions.

As shown in Fig.~\ref{fig:ratio-phi-cb}, we compare the medium modifications of the azimuthal angular correlations of $\gamma+$D and $\gamma+$B in $0-10\%$ Pb+Pb collisions at $\sqrt{s_{NN}}=$5.02 TeV. We find that the total modification of $\gamma+$B angular correlations is visibly smaller than that of $\gamma+$D at the back-to-back region~($\Delta \phi\sim \pi$), which may indicate that the in-medium $P_T$-broadening effect of bottom quark seems weaker than that of charm quark. However, we also observe that the enhancement of the ratio (PbPb/pp) of $\gamma$+B at $\Delta\phi<\pi/2$ is larger than that of $\gamma$+D, which may be related to the steeper initial distribution of $\Delta\phi_{\gamma B}$ at this region compared to that of $\Delta\phi_{\gamma D}$ as shown in Fig.~\ref{fig:gDB-pp}.

To estimate the net contributions of collisional (elastic) and radiative (inelastic) mechanisms to the total angular de-correlations of $\gamma+$HF, we also show the respective results plotted as the red and blue lines in Fig.~\ref{fig:ratio-phi-cb}. Note that the notations ``Rad. only'' and ``Coll. only'' represent the the medium modification of $\Delta\phi_{\gamma D}$ attributed to radiative and collisional energy loss as a part of the total energy loss respectively, while the parameters $\hat{q}$ and $\kappa$ are fixed.
 We find that the radiative mechanism seems to play more important role in the total medium modification of $\Delta\phi_{\gamma D}$. As for $\gamma+$B, the contributions from radiative and collisional mechanisms are comparable, while the contribution from radiative mechanism is obviously smaller than that of $\gamma+$D. Due to the so-called ``dead-cone'' effect, the medium-induced gluon radiation can be suppressed by the heavy quark mass, especially for the heavier bottom quarks compared to charm quarks, which leads to smaller medium modification shown in $\gamma+$B compared to $\gamma+$D at the back-to-back region ($\Delta\phi\sim\pi$).

To make a deep understanding to the $P_T$-broadening caused by the collisional and radiative energy loss mechanisms, we estimate the transverse momentum gained by heavy quarks after a certain propagation length (L=4 fm) in a static medium (T=400 MeV) versus initial heavy quark $p_T$, as shown in Fig.~8. We find that at $p_T<5$ GeV the collisional energy loss of charm quarks dominates the momentum broadening, but the radiative energy loss plays more important role in the kinematics of higher transverse momentum. As for bottom quarks, the crossover point of these two mechanisms appears at 10 GeV, because bottom quarks suffer stronger ``dead-cone'' effect due to their larger mass. Hence we suggest that the future measurements of the angular de-correlations of $\gamma+$D and $\gamma+$B at the LHC may be helpful to understand the mass effect of charm and bottom quarks in heavy-ion collisions from a new perspective differing from that of the parton energy loss.

\begin{figure}[!t]
\begin{center}
\vspace*{0.1in}
\includegraphics[width=3.4in,height=4.1in,angle=0]{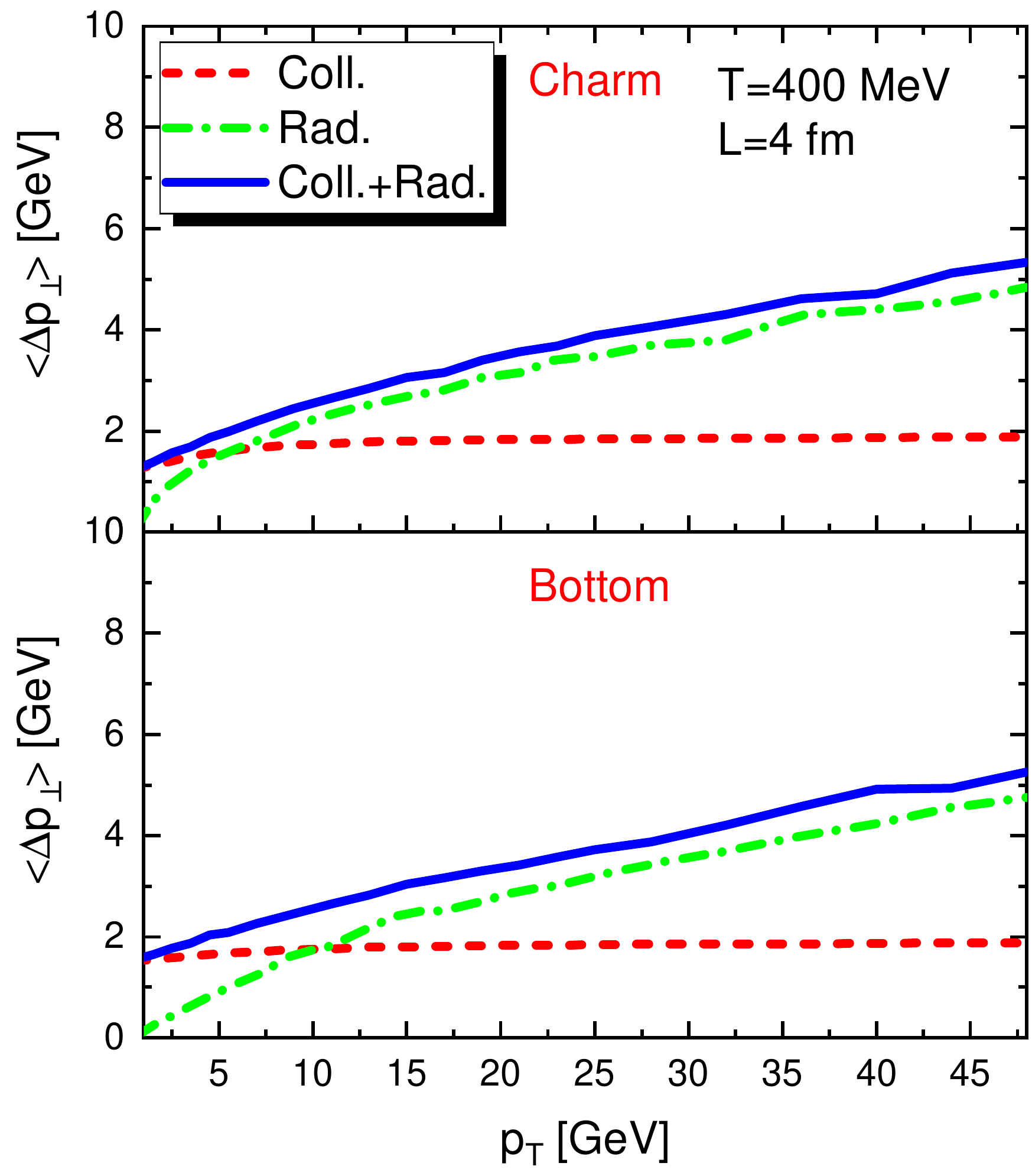}
\caption{The transverse momentum gained by charm (upper panel) and bottom (lower panel) quarks after a certain propagation length (L=4 fm) in a static medium (T=400 MeV) as a function of initial heavy quark $p_T$.}
\label{fig:pperp}
\end{center}
\end{figure}

\begin{figure}[!t]
\begin{center}
\vspace*{0.1in}
\includegraphics[width=3.4in,height=3.8in,angle=0]{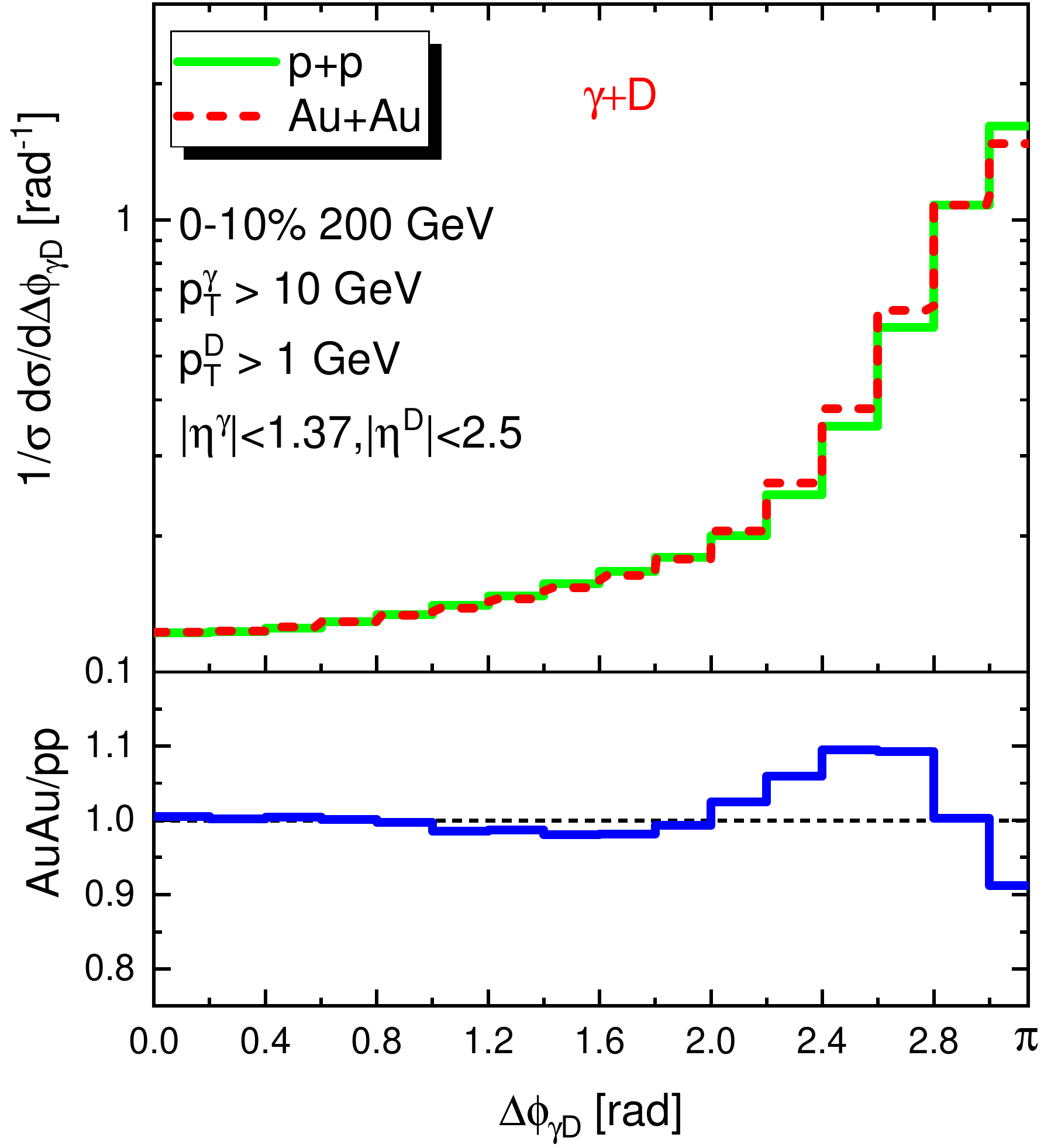}
\vspace*{0.1in}
\caption{The normalized distribution of the azimuthal angular difference ($\Delta \phi_{\gamma D}=|\phi_{\gamma}-\phi_{D}|$) between the isolated-photon and D meson in p+p and 0-10\% Au+Au collisions at $\sqrt{s_{NN}}$=200 GeV. The ratio of the normalized distributions in Au+Au to that in p+p is also plotted in the lower panel.}
\label{fig:ppAAphi200}
\end{center}
\end{figure}

Last but not least, we also present the predicted medium modification of $\gamma+$D angular correlations at the RHIC energy. In Fig.~\ref{fig:ppAAphi200}, we show the normalized $\Delta \phi_{\gamma D}$ distributions in p+p and central $0-10\%$ Au+Au collisions at $\sqrt{s_{NN}}$=200 GeV, as well as their ratio (PbPb/pp) in the lower panel. Since the QGP formed at the RHIC energy has lower average temperature than that at the LHC, the corresponding $P_T$-broadening of heavy quarks may be smaller. However, the final-state modification of $\gamma+$D angular correlation in nucleus-nucleus collisions also depends on the initial $\Delta \phi_{\gamma D}$ distribution. We find that the $\Delta \phi_{\gamma D}$ spectra in p+p collisions at RHIC energy is steeper than that at LHC energy (as shown in Fig.~\ref{fig:ppAAphi}), as a result, a visible modification on the $\Delta \phi_{\gamma D}$ distributions in Au+Au collisions can still be observed. We are looking forward to such measurements relating to $\gamma+$HF angular de-correlations can also be implemented both at the RHIC and the LHC energy.

\section{Summary}
\label{sec:sum}
In this paper, we present the first theoretical study of the azimuthal angular de-correlations of $\gamma+$HF in high-energy nuclear collisions as a new probe of the in-medium $P_T$-broadening of heavy quarks traversing the quark-gluon plasma. The p+p baseline is produced by the event generator SHERPA which computes the next-to-leading order matrix elements matched with parton shower effects. The in-medium heavy quark evolution is implemented by a Monte Carlo Langevin simulation, which takes into account the partonic elastic and inelastic interactions.

In Pb+Pb collisions at $\sqrt{s_{NN}}=$5.02 TeV, we find considerable suppression at $\Delta\phi_{\gamma D}\sim\pi$ and enhancement at $\Delta\phi_{\gamma D}<2.8$ in $\gamma+$D azimuthal angular distribution compared to the p+p baseline. The angular de-correlations between photon and D meson can be used to probe the in-medium $P_T$-broadening of charm quarks in experiment. By analysing the contributions from different kinematics of D meson in the $\gamma+$D angular distribution, we find that lower $p_T$ D meson play an important role to determine the overall modification patterns of the $\Delta\phi_{\gamma D}$. As for the estimation at higher $p_T^{\gamma}$ ranges, the angular de-correlations between photon and D meson are found to be not significant. Furthermore, to display the respective impact of energy loss and $p_T$ broadening on the final-state $\gamma$+D observable simultaneously, we construct the 2D correlations diagrams between $x_{J}^{\gamma D}$ and $\Delta\phi_{\gamma D}$ both in p+p and $0-10\%$ Pb+Pb collisions at $\sqrt{s_{NN}}=$5.02 TeV. We observe that $x_{J}^{\gamma D}$ shifts towards smaller value representing the energy loss of charm quarks, and the strengthen of ($x_{J}^{\gamma D}, \Delta\phi_{\gamma D}$) correlations at smaller $\Delta\phi_{\gamma D}$ region indicating the $P_T$-broadening of charm quarks relative to the direct photon.

Additionally, we also investigate the angular de-correlations of $\gamma+$B in Pb+Pb collisions at $\sqrt{s_{NN}}=$5.02 TeV. We observe weaker medium modifications of azimuthal angular correlation of $\gamma+$B compared to that of $\gamma+$D in Pb+Pb collisions at $\Delta \phi\sim \pi$. And we demonstrate that the difference mainly results from the medium-induced gluon radiation, which may be helpful to understand the ``dead-cone'' effect of charm and bottom quarks in heavy flavor physics. At last, we also present the calculated medium modification of $\gamma+$D angular correlations at the RHIC energy, and a visible angular de-correlation of $\gamma+$D in $0-10\%$ Au+Au collisions at $\sqrt{s_{NN}}=$200 GeV is predicted.

{\it Acknowledgments:}
 The authors would like to thank S. Chen, P. Ru, R. Ma for their helpful comments, and Frank Siegert for providing the Run card of SHERPA simulations. This research is supported by the Guangdong Major Project of Basic and Applied Basic Research No.~2020B0301030008, the Science and Technology Program of Guangzhou No.~2019050001 and Natural Science Foundation of China with Project Nos.~11935007 and 12035007. Sa Wang is also supported by China Postdoctoral Science Foundation under project No.~2021M701279.

\end{document}